\documentclass[aip,graphicx]{revtex4-1}

\usepackage{amssymb,amsmath,amsfonts}
\usepackage{graphicx,color}
\usepackage[english]{babel}
\usepackage{tabularx}
\usepackage{bm}
\usepackage{color,soul}
\usepackage{ulem}

\newcommand{\1}{\begin{equation}}
\newcommand{\2}{\end{equation}}
\newcommand{\ea}{\begin{eqnarray}} 
\newcommand{\ee}{\end{eqnarray}}
\newcommand{\4}[2]{{\frac{#1}{#2}}}
 
\newcommand{\Sum}[2]{{\sum\limits_{#1}^{#2}}}

\draft 

\begin{document}


\title{Light-controlled Assembly of Active Colloidal Molecules}

\author{Falko Schmidt}
\thanks{FS and BL contributed equally}
\affiliation{Department of Physics, University of Gothenburg, SE-41296 Gothenburg, Sweden}

\author{Benno Liebchen}
\thanks{FS and BL contributed equally}
\affiliation{Institut f\"{u}r Theoretische Physik II: Weiche Materie, Heinrich-Heine-Universit\"{a}t D\"{u}sseldorf, D-40225 D\"{u}sseldorf, Germany}

\author{Hartmut L\"owen} 
\email{hlowen@hhu.de}
\affiliation{Institut f\"{u}r Theoretische Physik II: Weiche Materie, Heinrich-Heine-Universit\"{a}t D\"{u}sseldorf, D-40225 D\"{u}sseldorf, Germany}

\author{Giovanni Volpe}
\email{giovanni.volpe@physics.gu.se}
\affiliation{Department of Physics, University of Gothenburg, SE-41296 Gothenburg, Sweden}

\date{November 02, 2018}

\begin{abstract}
Thanks to a constant energy input, active matter can self-assemble into phases with complex architectures and functionalities such as living clusters that dynamically form, reshape and break-up, which are forbidden in equilibrium materials by the entropy maximization  (or free energy minimization) principle. The challenge to control this active self-assembly has evoked widespread efforts typically hinging on engineering of the properties of individual motile constituents. Here, we provide a different route, where activity occurs as an emergent phenomenon only when individual building blocks bind together in a way that we control by laser light. Using experiments and simulations of two species of immotile microspheres, we exemplify this route by creating active molecules featuring a complex array of behaviors, becoming migrators, spinners and rotators. The possibility to control the dynamics of active self-assembly via light-controllable nonreciprocal interactions will inspire new approaches to understand living matter and to design active materials.
\end{abstract}


\maketitle

\section{Introduction}
One promising approach to create functional materials as required by 21st century's technologies is provided by self-assembly. Here, a basic starting point is to explore and control the binding of particles in a molecule, which works in principle on all scales, from atoms to colloids. As compared to their atomistic counterparts, the formation of colloidal molecules \cite{Manoharan2003, Bianchi2006, Wang2012, Nguyen2017, Niu2017, Niu2017a} offers an enhanced control, based on the possibility to design their shapes and coatings on demand, as illustrated e.g. by the admirable achievements on ``patchy colloids" \cite{Manoharan2003, Bianchi2006, Wang2012, Nguyen2017}.

While many works exploring the self-assembly of molecules focus on equilibrium systems, active particles that locally inject energy into a material open promising new horizons for self-assembly. These active particles are intrinsically away from thermal equilibrium \cite{Ramaswamy2010, Menzel2015, Zottl2016, Bechinger2016}, which allows them to conquer a new level of complexity. This new complexity finds its perhaps most spectacular expression in the hierarchical self-organization of biological matter, often leading to functionalities such as clustering \cite{peruani2006nonequilibrium}, navigation \cite{Dusenbery2009}, self-healing \cite{Trask2007}, or reproduction \cite{Narra2006}. It has been recently theoretically suggested \cite{Ivlev2015, Soto2014, Soto2015} that activity can also be exploited to form active molecules where the nonequilibrium settings allow the spontaneous emergence of new physical properties: the nonequilibrium chemical interaction between two immotile particles of different species will be generally non-reciprocal \cite{Ivlev2015}, resulting in the formation of colloidal molecules that may spontaneously acquire motility \cite{Soto2014, Soto2015}. This differs conceptually from active molecules involving components that are individually motile \cite{Popescu2011, Baraban2012, Zhang2016, Singh2017, ilday2017rich, Loewen2018}, and also from molecules that are first prepared with an irreversible bonding and then aquire motility as an independent additional effect in electric ac-fields \cite{Ma2015a, Ma2015b}. 

\begin{figure*}
\includegraphics[width=0.95\textwidth]{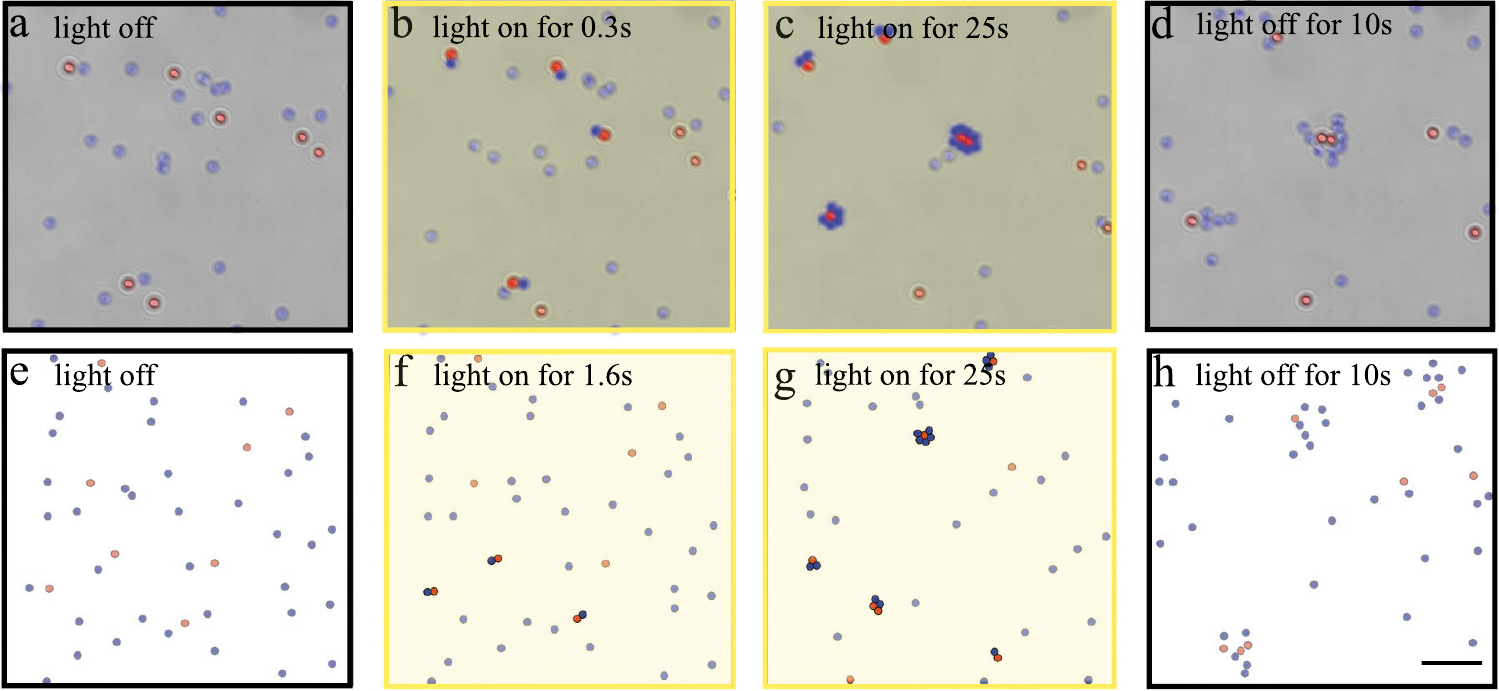}
\caption{\label{fig1}{\bf Spontaneous assembly of active colloidal molecules from immotile building blocks.} 
Series of snapshots from an experiment (top; supplemental video~1) and a simulation (bottom; supplemental video~2).
(a,e) Initially there are building blocks of two non-interacting species: light-absorbing (red, Fig.~\ref{fig2}a) and non-absorbing (blue, Fig.~\ref{fig2}b) 
spherical colloidal particles, which perform standard Brownian motion.
(b,f) When the sample is illuminated, the absorbing particles warm their surroundings and, when they randomly meet a non-absorbing particle, they join 
forming a self-propelling Janus dimer (Fig.~\ref{fig2}c).
(c,g) As time passes, the dimers collect additional particles and more complex structures emerge, which feature more complex behaviors, such as 
\emph{migrators} (Figs.~\ref{fig2}c-e), \emph{stators} (Fig.~\ref{fig2}f), \emph{spinners} (Fig.~\ref{fig2}g), and \emph{rotators} (Fig.~\ref{fig2}h).
(d,h) When the illumination is switched off, the active molecules disassemble and their component particles diffuse away.
The simulations use $1\,{\rm \mu m}$ and $1\,{\rm s}$ as length and time units, and values of the particle radii, diffusion constants, pair-molecule velocities as in the experiments; see the Methods for a detailed discussion. The scale bar is $10\,{\rm \mu m}$ and the laser intensity $I=0.08\,{\rm \mu W\mu m^{-2}}$.
}
\end{figure*}

Here, we experimentally demonstrate the formation of light-controllable active colloidal molecules from a suspension of light-absorbing and non-absorbing immotile microspheres immersed in a subcritical liquid mixture. When illuminated, the liquid surrounding the light-absorbing microparticles warms up. 
This temperature increase is isotropic and, therefore, does not lead to self-propulsion of the light-absorbing particles, differently from previous work on light-activated colloids \cite{Palacci2013}; however, it induces attraction among nearby colloids, which come together because of phoretic interactions at intermediate interparticle distances \cite{Palacci2013, liebchen2018interactions, yu2018chemical} (probably thermophoresis) and stick together because of short range atractions (probably Van der Waals and perhaps critical Casimir forces) \cite{Hertlein2008, Paladugu2016, Nguyen2016}.
For example, a Janus dimer is formed when this light-induced attraction holds a non-absorbing microsphere and an absorbing one together. This dimer experiences a temperature gradient and moves by phoretically \cite{Anderson1989, Golestanian2005, Volpe2011, Buttinoni2012, Kummel2013, Wuerger2015, Samin2015}. When more microspheres come together, they form more complex molecules including stators, migrators, spinners and rotators. Importantly, we remark that the emergence of directed motion of two binding immotile particles is not an obvious consequence of symmetry breaking -- rather, it is forbidden in equilibrium on the relevant scales, and crucially exploits the presence of a nonequilibrium environment. 
The striking new feature of the present approach is that motility occurs as an emergent nonequilibrium phenomenon from particle interactions that are controllable by light. This establishes a generic route to control nonreciprocal interactions among colloids by light, which is based on the laser-stimulated production of a phoretic field (e.g. chemicals, temperature, ions) by one species that attracts another species without causing a counteraction. This route offers control of the dynamics of active molecule formation, which can be switched on, off, paused or resumed on demand, and can be used to statistically control the composition of the system e.g. with respect to the ratio of linear and chiral swimmers, as we demonstrate below. 
Remarkably, also the concept of emergent motility itself may inspire new collective phenomena beyond those featured by individually motile particles \cite{Theurkauff2012, Palacci2013, Buttinoni2013, Colberg2015, Klapp2016, Ebbens2016, Marchetti2016}.


\begin{figure}
\includegraphics[width=0.5\textwidth]{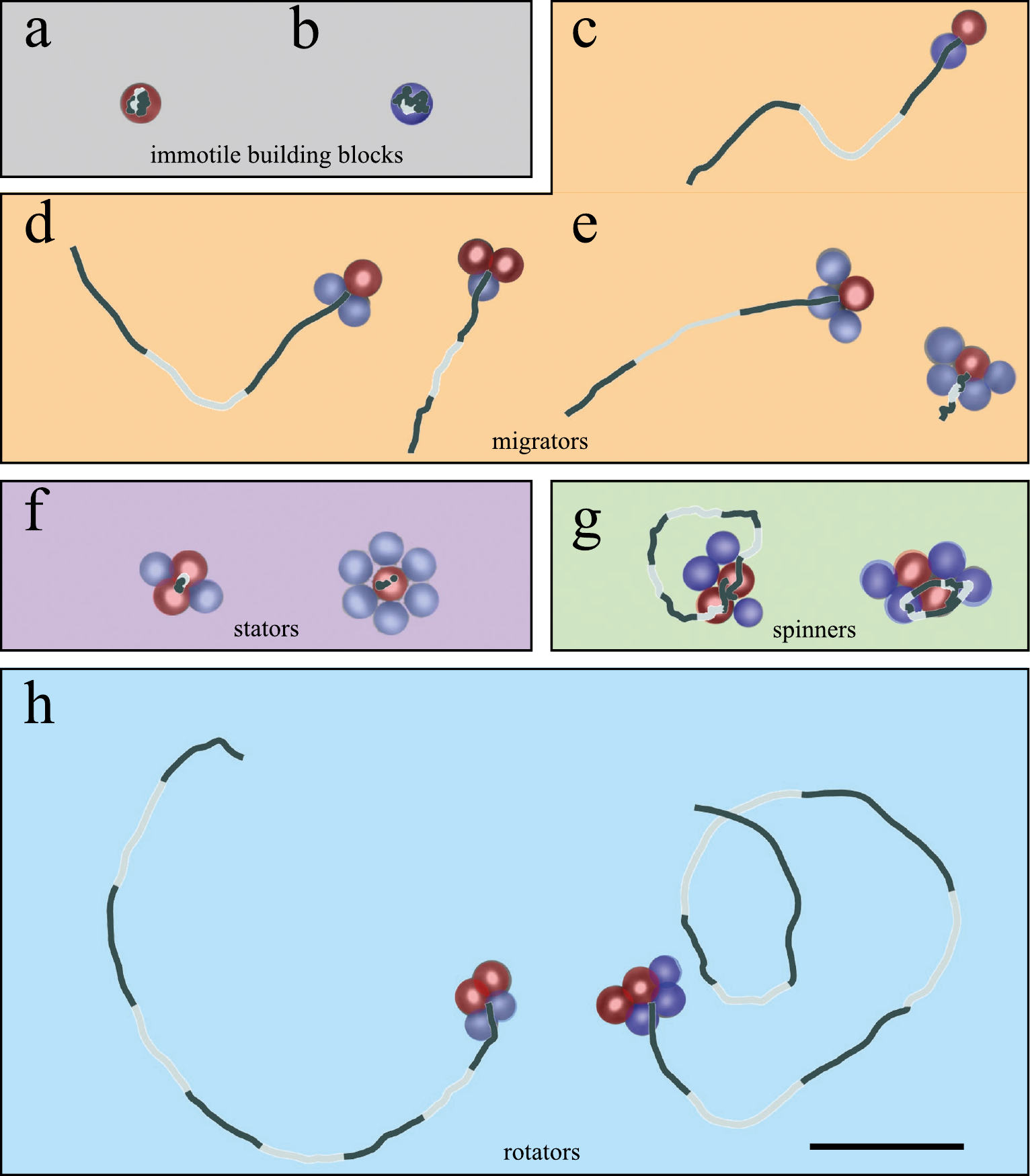}
\caption{\label{fig2}{\bf The zoologic garden of active molecules.}
While the constituting building blocks are immotile symmetric spherical (a) light-absorbing and (b) non-absorbing particles, the emerging active molecules present more and more complex shapes and behaviors as time evolves.
We observe: 
(c) dimers, (d) trimers, and (e) more complex axis-symmetric active molecules, which perform linear active Brownian motion (\emph{migrators});
(f) highly symmetric molecules that do not show activity (\emph{stators});
(g) rotationally-symmetric chiral active molecules which rotate almost without translating (\emph{spinners});
(h) asymmetric chiral active molecules which swim in circles (\emph{rotators}).
In all cases, lines represent particle trajectories; grey and black segments correspond to 1-s stretches.
The scale bar is $5\,{\rm \mu m}$.
}
\end{figure}

\section{Results}

\subsection{Experiments}

\begin{figure*}
\includegraphics[width=1.0\textwidth]{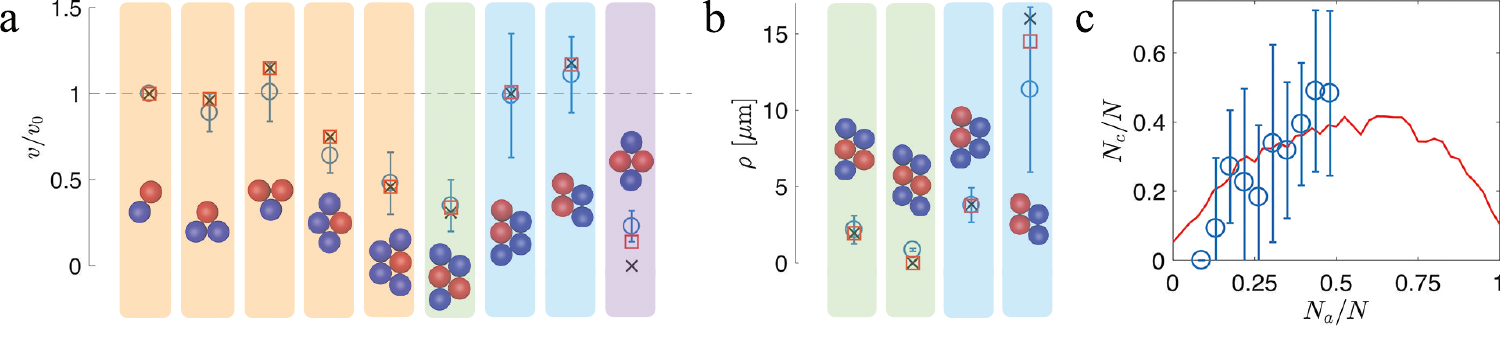}
\caption{\label{fig3}{\bf Quantitative agreement between experiments, simulations and analytical predictions.}
(a) The speed $v$ (normalized by the speed of the dimer $v_0$) and (b) swimming radius $\rho$ of active molecules obtained experimentally from the measured trajectories (blue circles and relative error bars representing a standard deviation), numerically from the simulated trajectories (red squares), and analytically for $D=0$ from Eqs.~\ref{eq1} and \ref{eq2} (black crosses) are all in good agreement.
(c) Chirality control: The fraction of chiral active molecules, ($N_{\rm c}/N$), can be controlled by the fraction of absorbing particles ($N_{\rm a}/N$) in the initial suspension. The blue circles and relative error bars representing a standard deviation are the experimental data, and the red solid line represents the simulation results. 
Simulation parameters in the Methods.
}
\end{figure*}

We consider a mixture of light-absorbing and non-absorbing colloidal particles (radius $R\approx 0.49\,{\rm \mu m}$) made of silica with and without iron-oxide inclusions in a near-critical water-lutidine mixture in a quasi two-dimensional sample chamber (see Methods for details). 
Both species of particles settle down above the bottom wall of the sample at approximatively the same distance, because their specific density and electrostatic interactions are similar; when compared with the same particles in the bulk of the solution, this increases the viscous drag acting on the particles by a constant factor \cite{happel2012low}, but does not otherwise qualitatively alter their dynamics.
Without illumination, both species show Brownian diffusion (Fig.~\ref{fig1}a), as can also be seen from their trajectories (Figs.~\ref{fig2}a and \ref{fig2}b). However, when illuminated, the fluid heats up only in the vicinity of the light-absorbing particles so that the fluid locally demixes and induces attractive interactions with other particles in the vicinity. 
The strength of the attractive interactions increases when enhancing the laser power, while control experiments in water show that no molecules form.
When an absorbing microsphere comes close to a non-absorbing one, we observe the formation of a heterogeneous dimer (Fig.~\ref{fig1}b), which, unlike the colloidal building blocks it consits of, starts to move ballistically. 
This ballistic motion is forbidden in equilibrium and is therefore not a simple consequence of symmetry breaking, but also involves nonequilibrium fluctuations.
In analogy to self-propelled Janus colloids, we call the emerging dimer a 
\emph{Janus dimer}.
Janus dimers represents the simplest active molecules. 
Their speed and rotational diffusion amounts to $v_2=2.0\pm 0.4\,{\rm \mu m\,s^{-1}}$ and $D_r=0.11\pm 0.05\,{\rm s^{-1}}$.
In the course of our experiments, the size of the molecules keeps on growing as time passes and clusters coalesce: the dimers move around and collect additional particles so that more complex structures emerge, which feature more complex behaviors, as shown in Fig.~\ref{fig1}c. 
These behaviors are intimately linked to the symmetry properties of the resulting active molecules. 
There are axis-symmetric molecules that behave as \emph{migrators} performing linear active Brownian motion, such as dimers (Fig.~\ref{fig2}c), trimers (Fig.~\ref{fig2}d), as well as larger structures (Fig.~\ref{fig2}e), which move with the absorbing microsphere in front 
\cite{Volpe2011, Buttinoni2012, Kummel2013, Wuerger2015, Samin2015}. More symmetric shapes perform standard Brownian motion, behaving as \emph{stators} (i.e. passive colloidal molecules, Fig.~\ref{fig2}f). Finally, chiral shapes behave as either \emph{spinners} (Fig.~\ref{fig2}g) or \emph{rotators} (Fig.~\ref{fig2}h), featuring different forms of chiral active Brownian motion. Importantly, the assembly mechanism of these active molecules is fully reversible and they melt by thermal diffusion when the light is switched off (Fig.~\ref{fig1}d).
The speed and rotation frequency of these molecules depend on the details of their structure and composition, as shown in Figs.~\ref{fig3}a and \ref{fig3}b. However, for very large molecules, where the phoretic drift contributions of the absorbing--non-absorbing pair are distributed randomly within a molecule, they scale with $N^{-1/2}$, where $N$ is the number of monomers in the molecule.
All molecules generally attract each other. For the laser power used in Fig.~\ref{fig1}, these interactions are strong enough to bind colliding molecules together. However, for appropriate weaker attractions (lower laser power), dimers do not grow towards larger molecules allowing to generate a gas containing only monomers and dimers.

\begin{figure}
\includegraphics[width=0.5\textwidth]{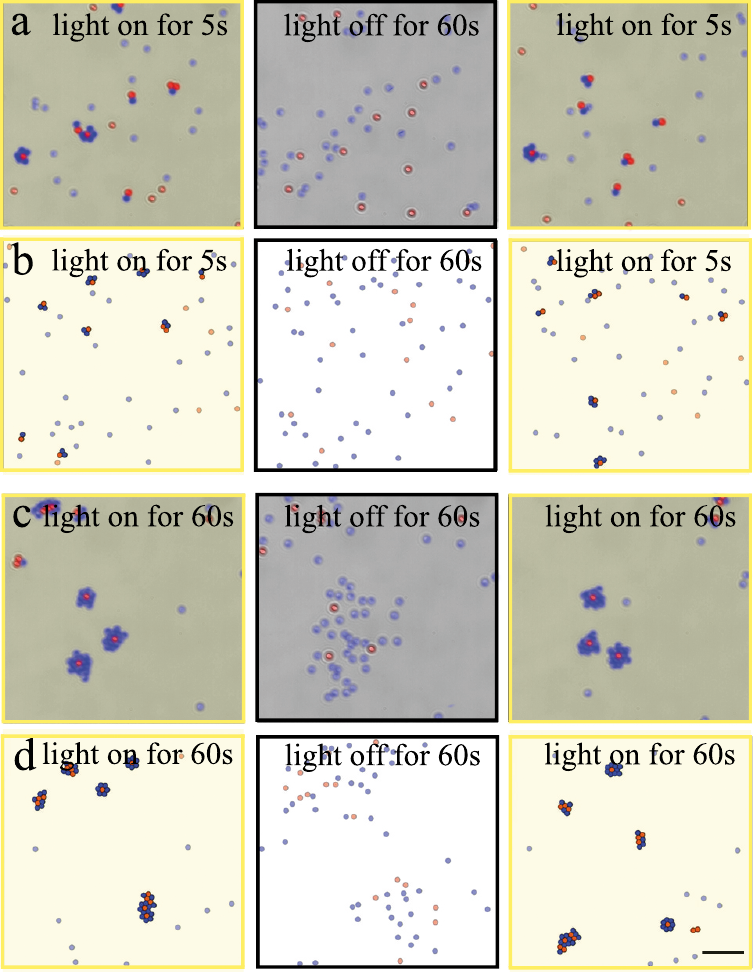}
\caption{\label{fig4}{\bf Controlling the assembly of active molecules by periodic illumination.}
By subjecting the system to periodic illumination such that the light is periodically on for (a,b) $5\,{\rm s}$ and (c,d) $60\,{\rm s}$ and off always for $60\,{\rm s}$, 
it is possible to assemble metastable active colloidal molecules with a controllable size, shape and functionality, 
as their characteristic size increases with the light-on period. 
Each row features a series of three (a,c) experimental and (b,d) simulation snapshots taken at the end of a series of light-on, light-off, light-on periods.
{The corresponding movies are supplemental videos (a) 3, (b) 4, (c) 5, and (d) 6.}
Simulation parameters as in Fig.~\ref{fig1} and provided in the Methods.
The scale bar is $10\,{\rm \mu m}$.
}
\end{figure}

\subsection{Theoretical model}

To identify the key ingredients determining the emergence and dynamics of the colloidal molecules, we develop a phenomenological model based on the interplay of attractive interactions, steric repulsions and particle diffusion inducing molecule formation, and the presence of nonreciprocal phoretic interactions creating self-propulsion (see Methods for details and equations).
The model includes a mixture of light-absorbing and non-absorbing overdamped Brownian particles with radii $R = 0.49\,{\rm \mu m}$ following overdamped Langevin dynamics including Gaussian white noise, which represents monomer diffusion and automatically determines translational and rotational diffusion of the emerging molecules. Light-absorbing particles interact attractively with all particles in their vicinity. These attractions involve phoretic interactions, which are effectively screened on the scale of a few particle radii \cite{liebchen2018interactions}, and also shorter-ranged attractions (van der Waals and perhaps critical Casimir interactions), which bind the particles in a molecule almost rigidly (note that unscreened phoretic interactions decaying quadratically with distance, would lead to a rapid collapse of the system). Details of the overall attractions are unimportant for speed, rotation rates and shapes of the emerging molecules and can be modeled phenomenologically e.g. using Lennard-Jones interactions. Non-absorbing particles repel each other sterically (Weeks-Chandler Anderson repulsions), but in the vicinity of an absorbing particle heating up the solvent locally, they weakly attract each other, which we include in the model using Lennard-Jones interactions with a comparatively small attraction depth. Accordingly, as in the experiments, non-absorbing particles stay together within a molecule, but rarely in bulk.
To model directed motion of pairs of absorbing and non-absorbing particles, we pursue the following physical picture: Light-absorbing particles act as sources of phoretic fields (probably temperature) in the near-critical binary liquid \cite{Buttinoni2013, Wuerger2015, Samin2015}. The gradients of these fields create a flow in the interfacial layer of the non-absorbing colloids leading to directed motion (phoresis) of the non-absorbing particles towards the absorbing ones \cite{liebchen2018interactions}. Thus, when close to each other, the non-absorbing  particles effectively push the absorbing ones forward, leading to directed motion of absorbing--non-absorbing pairs, similarly to the case of Janus particles \cite{Buttinoni2013, Wuerger2015, Samin2015}. In addition, the non-absorbing particle displaces the relevant phoretic fields so that the absorbing particle itself experiences a phoretic gradient leading to phoretic motion. We model these effects by assigning an effective nonreciprocal force to absorbing--non-absorbing pairs. When several non-absorbing particles attach to an absorbing one, they collectively push it forward, i.e. their  contributions to the translation velocity of the molecule superimpose. As a minor correction to this picture, we account for the fact that non-absorbing particles attaching to a molecule containing non-absorbing particles see a displaced phoretic field; thus, non-absorbing particles attaching adjacently to an absorbing particle mutually reduce their contributions to the molecules' propulsion slightly.
This model applies to rather small molecules and also to large molecules containing a low fraction of absorbing particles; for a high density of absorbing particles, in the experiments the solvent heats up over an extended spatial region leading to a breakdown of the short-range attractions and to the emergence of repulsive forces between the absorbing and the non-absorbing particles.

Simulating this model, we observe the formation of active molecules as in the experiments, as shown by Figs.~\ref{fig1}e-h: from a primordial broth of immotile single particles (Fig.~\ref{fig1}e), first dimers form (Fig.~\ref{fig1}f) and later grow into more complex active molecules (Fig.~\ref{fig1}g). Switching off the attractive interactions (corresponding to switching off the illumination in the experiment), the active molecules disassemble and their constituent particles diffuse away from each other (Fig.~\ref{fig1}h).

Besides this qualitative agreement, our model allows for the prediction of the complete table of emerging active molecules and their properties (i.e. speed, chirality and swimming radius) in quantitative agreement with the experiment (see Methods for expressions and details of the calculation),
as can be seen in Fig.~\ref{fig3}. 
We discretize the simulated trajectory at $0.1\,{\rm s}$ ($10\,{\rm Hz}$) and calculate speeds and for circle swimmers also swimming radii (i.e. speed over rotation frequency) using finite differences as for the experimental data. The blue circles in Fig.~\ref{fig3}a represent experimentally measured speeds of various active molecules relative to the dimer speed, whereas red squares show speed values found in simulations, which are in quantitative agreement with the experimental results. Blue circles in Fig.~\ref{fig3}b represent the experimentally measured swimming radii $\rho$ of various chiral active molecules ($\rho = \infty$ for achiral molecules), whereas the red squares represent the corresponding swimming radii obtained from the simulations, which also agree quantitatively with the experimental results. Importantly, the specific choice of simulation parameters (in particular, the values of the cut-offs in the interaction potentials) has little bearing on the set of emerging active molecules or on their speeds, chiralities and swimming radii, which supports the validity of the model.
While the set of emerging molecules can only be predicted by simulations, their speeds and rotation radii can also be predicted analytically as we discuss in the Methods. These predictions are represented by the black crosses in Figs.~\ref{fig3}a and \ref{fig3}b and are 
in quantitative agreement with the experiments and simulations.

\subsection{Controlling the clusters' behavior and functionality}

The emergent behavior and functionality of this system can be controlled through a wealth of parameters. For example, in Fig.~\ref{fig3}c we show that it is possible to change the fraction of achiral (e.g. Figs.~\ref{fig2}c-e) versus chiral (e.g. Figs.~\ref{fig2}g-h) active molecules by changing the relative abundance of absorbing versus non-absorbing particles present in the initial suspension. The relative abundance of chiral active molecules increases with increasing deviation of the absorbing particle fraction from both 0 and 1. We can understand this result considering that only molecules featuring a common symmetry axis of composition and shape (Figs.~\ref{fig2}c-e) swim linearly, while molecules not featuring such a symmetry (Figs.~\ref{fig2}f-g) are generally chiral (exceptions are molecules where rotational torques balance unsystematically); while such a symmetry is generally present for molecules consisting of only two or three individual colloids, all active molecules that break it involve at least two absorbing and two non-absorbing spheres. This latter requirement for being chiral is fulfilled for more molecules when the ratio of absorbing and non-absorbing colloids in the initial ensemble approaches 1/2. Also in this case, we find a good agreement between our experimental results (blue symbols in Fig.~\ref{fig3}c) and the results of the simulations (red solid line). 
In fact, the simulations permit us to go beyond what can be realized experimentally as in experiments large assemblies including many absorbing colloids increase the temperature well beyond the critical point, inducing strong demixing around the colloids that prevents the formation of stable molecules involving attractive interactions. Also in this case, our simulations confirm the expectation described above, that the fraction of chiral to non-chiral molecules increases with increasing deviation of the fraction of absorbing to non-absorbing colloids from 1.

We can control the formation and growth of the molecules by tuning the light illumination. For example, it is possible to inhibit the formation of molecules larger than dimers by reducing the illumination to the minimum value where attraction between absorbing and non-absorbing monomers still occurs: Monomers can easily bind to each other but not to dimers; when occasionally trimers emerge, they typically decay on timescales of a few seconds.
For higher levels of illumination, the active molecules keep on growing in size indefinitely, if the illumination is constantly switched on. Despite this fact, we can control their size, shape and functionality by subjecting them to periodic illumination. This is interesting because many natural phenomena are subject to periodic excitation (e.g., circadian rhythms, molecular clocks) and can be exploited to engineer artificial systems (e.g., autonomous nanorobots). For example, we can consider a periodic pattern of illumination where the illumination is alternatively switched on and off. Figure~\ref{fig4}a shows that the active molecules that assemble when the light is periodically on for $5\,{\rm s}$ and off for $60\,{\rm s}$ are predominantly dimers and trimers, with few larger active molecules; this is in good agreement with simulations (Fig.~\ref{fig4}b). As the illumination time increases, the active molecules that form become larger and more complex, as can be seen in Fig.~\ref{fig4}c, where the light is periodically on for $60\,{\rm s}$ and off for $60\,{\rm s}$; this is also in agreement with simulations (Fig.~\ref{fig4}d). 

\section{Discussion}

We have demonstrated the light-controlled assembly of active colloidal molecules starting from a mixture of different species of immotile building blocks.
These molecules spontaneously acquire motility through non-reciprocal interactions of their immotile components and represent a new route to create active matter. Our proof-of-principle setup serves as a construction kit to assemble modular linear swimmers, migrators, spinners and rotators with light-controllable shape, size, speed and chirality. The table of the emerging active molecules and their characteristic properties can be quantitatively predicted by an effective model, which can be used in the future to design molecules and to predict their response to external fields, their large-scale collective behavior, and the properties of large molecules; the model may also be extended towards a more microscopic description \cite{Wuerger2015, Samin2015}, and to account for phoretic \cite{Saha2014, Pohl2014, Liebchen2015, Liebchen2017} and hydrodynamic interactions \cite{Zottl2016, Takatori2016}.
It will also be interesting to further explore the microscopic processes underlying the structure formation, explaining further how the structures are formed, how they can be controlled externally (e.g. by using spatiotemporal light modulators) and how individual molecules interact with each other.
The exemplified route to create activity from immotile building blocks serves as a new design principle 
for active self-assembly. This might be useful both from a materials perspective and to explore and design functionality in highly controllable synthetic systems, which is so far often restricted to uncontrollable biological environments.

\begin{acknowledgments}
We thank C. Lozano for useful discussions. This work was partially supported by the ERC Starting Grant ComplexSwimmers (grant number 677511), by Vetenskapsr{\aa}det (grant number 2016-03523), and by the German Research Foundation DFG within LO 418|19-1.
\end{acknowledgments}

\appendix
\section{Experimental setup}

We consider a suspension of colloidal particles in a critical mixture of water and 2,6-lutidine at the critical lutidine mass fraction $c_{\rm c} = 0.286$ with a lower critical point at the temperature $T_{\rm c} \approx 34^\circ{\rm C}$ \cite{Grattoni1993} (see Supplementary Fig.~S1).
The light-absorbing species consists of silica microspheres with absorbing iron-oxide inclusions (Microparticles GmbH), 
while the non-absorbing species consists of equally-sized plain silica microspheres (Microparticles GmbH). 
Both particle species have the same radius ($R = 0.49\pm0.03\,{\rm \mu m}$) and similar density ($\rho\approx 2\,{\rm g\, cm^{-3}}$).
The suspension is confined in a quasi-two-dimensional sample chamber realized between a microscope slide and a coverslip, where the particles sediment due to gravity.

A schematic of the setup is shown in Supplementary Fig.~S2. The motion of the particles is captured by digital video microscopy at $20\,{\rm fps}$.  Using a two-stage feedback temperature controller \cite{Paladugu2016, schmidt2018microscopic}, the temperature of the sample is adjusted to $T_0 = 31^\circ{\rm C}$, which is below $T_{\rm c}$ so that water and 2,6-lutidine are homogeneously mixed. In these conditions, the microspheres of both species behave as independent immotile Brownian particles and undergo standard diffusion (Fig.~\ref{fig1}a). To illuminate the sample we use a laser with a wavelength $\lambda = 532\,{\rm nm}$ 
at an intensity $I = 0.08\,{\rm \mu W\,\mu m^{-2}}$.
The increase of temperature in the vicinity of the light-absorbing particles is rather small ($\Delta T\approx 4^\circ{\rm C}$) so that they still diffuse as normal (non-active) Brownian particles. This is reflected also by the trajectories for the light-absorbing and non-absorbing particles in Figs.~\ref{fig2}a and \ref{fig2}b. 

\section{Details of the model}

The key ingredients of our model are simple as the molecule formation emerges from the interplay of attractive interactions, repulsive excluded volume interactions and particle diffusion. Additional nonreciprocal interactions based on (thermo-)phoretic interactions among heterogeneous colloids lead to a directed propulsion of these molecules.

We now specifically define the model whose key aspects we have described in the main text. To describe the dynamics of non-absorbing particles we employ the following overdamped Langevin equation \cite{volpe2014simulation}
\begin{equation}
\dot {\bf r}_i = -\4{1}{\gamma} \sum\limits_{j=1}^{N_{\rm a}}\nabla_{{\bf r}_i} V_1({\bf r}_i) 
-\4{1}{\gamma} \sum\limits_{j=1; j\neq i}^{N_{\rm p}} \nabla_{{\bf r}_i}V_2({\bf r}_i)
+ \sqrt{2D} {\bm \eta}_i(t) \; ,
\end{equation}
where the left sum extends over all $N_{\rm a}$ absorbing particles and the right one over all $N_{\rm p}$ non-absorbing particles (excluding particle $i$), $\gamma$ is the Stokes drag coefficient (assumed to be independent from the distance to other particles), $D$ is the Brownian diffusion coefficient and ${\bm \eta}_i$ represents Gaussian white noise with zero mean and unit variance.

The pair-interaction potential $V_1$ represents Lennard-Jones interactions acting among absorbing particles and between absorbing and non-absorbing particles. This interaction phenomenologically models a combination of different interactions in the system (phoretic attractions sometimes featuring an effective screening \cite{liebchen2018interactions}, short-ranged attractions and steric repulsions):
\begin{equation}
V_1({\bf r}_{ij})= 4\epsilon \left[\left(\4{\sigma}{r_{ij}}\right)^{12}-\left(\4{\sigma}{r_{ij}}\right)^6\right] \; ,
\end{equation}
where we have used $r_{ij}=|{\bf r}_{ij}|$, ${\bf r}_{ij}={\bf r}_i-{\bf r}_j$ and $\epsilon$ is the depth of the potential which crosses zero at $r=\sigma=2R/\epsilon^{1/6}$. Note that the precise form of the interactions does not affect the set of emerging molecules or their speeds and rotation rates; attractive Yukawa interactions in combination with Weeks-Chandler Anderson repulsions basically lead to the same results. In our simulations, we choose a cut-off distance of $8R$ for the Lennard-Jones interactions; also this choice does not affect the molecules and their speeds and hardly affects the kinetics of molecule formation.

Conversely to interactions among pairs involving an absorbing particle, interactions among non-absorbing particles are purely repulsive. However, when an absorbing particle is in the vicinity of one of the non-absorbing particles (we phenomenologically choose a critical distance of $r_{\rm c}=8R$) it heats up the solvent locally, leading to relatively weak attractions among the colloids. We therefore model the interactions among the non-absorbing particles as:
\begin{equation}
\nabla_{{\bf r}_i} V_2({\bf r}_{ij}) = 
\begin{cases}
\nabla_{{\bf r}_i} V_1({\bf r}_{ij}) & r_{ij} \leq 2R   \\
\alpha \nabla_{{\bf r}_i} V_1({\bf r}_{ij}) & r_{ij} > 2R\;\;\text{and} \\
&\Sum{k=1}{N_{\rm a}} \theta\left(r_{\rm c}-{\rm max}({r_{ik},r_{jk}}) \right)> 0 \\
0 & {\rm otherwise}  \end{cases} 
\end{equation}
where $\alpha \ll 1$ determines the relative interaction strength among non-absorbing particles compared to the interaction strength among absorbing particle. The key effect of the attractions among non-absorbing colloids in the vicinity of an absorbing one is that they tend to stay next to each other within a molecule, rather than moving almost freely along the rim of an absorbing particle.

When the laser is switched off, all particles are non-absorbing; however, when the laser is switched on, we describe the dynamics of absorbing particles by the following Langevin equation: 
\begin{equation}
\dot {\bf r}_i = -\4{1}{\gamma}\Sum{j=1; j\neq i}{N_{\rm a}} \nabla_{{\bf r}_i} V_1({\bf r}_i) + 
\Sum{j=1}{N_{\rm p}} {\bf v}_{ij} + \sqrt{2D}{\bm \eta}_i(t) \; ,
\end{equation}
where we have introduced the phoretic drift velocity ${\bf v}_{ij} = \nu_{ij} \theta(r_{\rm p}-r_{ij}){\bf r}_{ij}/r^3_{ij} $ describing directed motion of particle $i$ due to particle $j$. Physically, such a directed motion probably occurs because the (heat, chemical) gradients in the phoretic fields that the absorbing particles produce induce a stress in the interfacial layers of the non-absorbing particles leading to a localized solvent flow across their surface; this flow induces a directed motion of the non-absorbing particles towards the absorbing ones. When in close contact, the non-absorbing particles should push the absorbing particles forward; at the same time they probably displace the phoretic fields produced by the absorbing particles, which should induce a phoretic motion of the absorbing particles themselves. Both effects lead to a directed motion of the absorbing--non-absorbing pairs. For simplicity, we have implemented this propulsion in the equations of motion of the absorbing particles (implementing them in the equations of motion of the non-absorbing particles slightly changes the propulsion speed and rotation velocities of some of the molecules, but does not alter their qualitative behavior). Since directed motion sets in smoothly in the experiment, slightly before the individual molecules touch each other, we have used a $1/r^2$-scaling for the phoretic speed; this scaling is inspired by the corresponding far-field scaling of the  gradients in an unscreened phoretic field \cite{liebchen2018interactions}. The unit-step function $\theta$ defines a hard cut-off of the phoretic motion at a pair distance of $r_{\rm p}=2.8R$, roughly representing effective screening of the involved phoretic fields \cite{liebchen2018interactions}. The precise value of this cut-off does not affect the shapes of the emerging molecules or their speeds, and hardly affects the kinetics of molecule formation. Finally, we use the following expression for the coefficient of the phoretic velocity 
\begin{equation}\label{eq:neighbors}
\nu_{ij}=\nu_0 \left[1-\4{1}{6}\Sum{k=1, k\neq j}{N_{\rm p}}\theta(r_{\rm c}-r_{jk})\right] \; ,
\end{equation}
where $\nu_0$ determines the propulsion speed of an isolated absorbing--non-absorbing pair as $4R^2 \nu_0$. The term in square brackets accounts for the fact that adjacent non-absorbing particles attached to an absorbing one do not contribute fully independently to the propulsion-speed but mutually suppress their contributions slightly. This is partly caused by the fact that each non-absorbing particle displaces the phoretic field produced by an absorbing particle, so that each additional non-absorbing particle attaching to an absorbing one sees a different phoretic field. This may be viewed as a mutual shielding of a part of the absorbing colloid's surface. Here, we phenomenologically assume that each non-absorbing particle covers an angle of $\theta=\pi/2$ (2D projection) of the surface of the absorbing colloids (see Supplementary Fig.~3a). Thus, when two non-absorbing colloids adjacently attach to an absorbing colloid, the areas they cover overlap by an angle of $\theta_0=\pi/6$ (see Supplementary Fig.~3b); thus the net angle covered by each of the two non-absorbing colloids is $\pi/2-(\pi/6)/2=5\pi/6$. This reduces the contribution of each of them to the effective propulsion force from $\nu_0$ to $\nu_0 (5\pi/2)/(\pi/2)=(5/6)\nu_0$. Analogously, the contribution of a non-absorbing colloid in between two other ones is reduced by a factor of $1/3$. This is represented by the sum in Eq.~(\ref{eq:neighbors}), where $r_{\rm c}$ is an arbitrary value which must be chosen slightly larger than $2R$ (here $r_{\rm c}=2.3R$).

\subsection{Choice of parameters}

Here we provide the parameters we used in the implementation of the simulations. Nevertheless, we remark that the qualitative behavior of the colloidal molecules are very robust to the choice of parameters.
To allow for a straightforward comparison with experiments, we use $1\, {\rm \mu m}$ and $1\, {\rm s}$ as length and time units in all the simulations.
For Figs.~\ref{fig1} and \ref{fig4}, we have used $R=0.49\, {\rm \mu m}$ and $D=0.1\, {\rm \mu m^2 s^{-1}}$. We have further chosen $\epsilon/\gamma=10$, i.e. $\epsilon/(kT)=100$, so that the molecules, once formed, are robust against thermal fluctuations, as in the experiments. We use $\nu_0=24 R^2$ to fix the propulsion speed of a Janus dimer (absorbing and non-absorbing colloid) at contact distance of $2R$, which gives a value similar to that in the experiments. Finally, choosing $\alpha=1/20$ yields a comparatively weak mutual attraction among non-absorbing particles (strength $\epsilon/(kT)=5$), if an absorbing one is close by.
For Fig.~\ref{fig3}, which shows the molecule speed normalized by the pair speed, the absolute value of the propulsion speed is unimportant; here we have chosen $\nu_0=60R^2$ for efficiency of the simulations. We have further used $R=0.49\, {\rm \mu m}$, $\epsilon/\gamma=500, \alpha=1$ to strongly avoid fluctuations of the molecule shapes in the course of a simulation. Finally, we have used a somewhat stronger diffusion $D=0.41\, {\rm \mu m^2 s^{-1}}$ (Figs.~\ref{fig3}a and \ref{fig3}b) and $D=0.195\, {\rm \mu m^2 s^{-1}}$ (Fig.~\ref{fig3}c) to accelerate molecule formation and the corresponding convergence of ensemble averages; here, noise is largely negligible (for the used sampling rate) as can be seen from the comparisons with the noise-free analytical calculations of the relative speed and rotation rates of the molecules which are practically identical to the simulated ones (Figs.~\ref{fig3}a and \ref{fig3}c).

\subsection{Analytical predictions of speed and rotation frequency of molecules}

The speed, rotation radius and reorientation frequency of a given molecule can be calculated analytically in the zero-noise limit by assuming hexagonal close packing within the molecule. Consider the molecule as a rigid body and the following balance conditions for the effective forces ${\bf F}_i:=\gamma \Sum{j=1}{N_{\rm p}}v_{ij}$ (${\bf F}_i=0$ for non-absorbing particles) and the associated effective torques
\begin{equation}
\Sum{i=1}{N} {\bf F}_i -\gamma \dot {\bf r}_i=0; \quad \Sum{i=1}{N} ({\bf r}_i-{\bf R})\times ({\bf F}_i-\gamma \dot {\bf r}_i)=0 \; ,
\end{equation}
where we sum over all $N$ particles in a molecule and choose ${\bf R}$ as the centre of mass of the molecule.
From here, using polar coordinates, we readily find the considered molecule's velocity and rotation frequency as 
\begin{equation}\label{eq1}
{\bf v} = \dot {\bf R}={\frac{1} {N \gamma}}\Sum{i=1}{N}{\bf F}_i
\end{equation}
and 
\begin{equation}\label{eq2}
\omega={1 \over \gamma}\left| {\Sum{i=1}{N}{{\bf x}_i \times {\bf F}_i} \over \Sum{i=1}{N} {\bf x}_i^2} \right| \; ,
\end{equation}
where ${\bf x}_i$ is the relative coordinate of sphere $i$ with respect to the centre of mass of the active molecule. From Eq.~(\ref{eq2}) the gyration radius $\rho$ follows as $\rho={|{\bf v}| \over \omega}$.
In Supplementary Fig.~3c, we exemplarily illustrate the analytical calculation of the swimming speed and swimming radius for  a molecule consisting of 5 particles (2 absorbing and 3 non-absorbing).


\end{document}